\documentclass[12pt]{article}
\usepackage{graphicx}
\def\beq{\begin{equation}}
\def\eeq{\end{equation}}
\def\bea{\begin{eqnarray}}
\def\eea{\end{eqnarray}}

\def\roughly#1{\mathrel{\raise.3ex\hbox
{$#1$\kern-.75em\lower1ex\hbox{$\sim$}}}}

\def\sss{\scriptscriptstyle}

\def\bd{B_d^0}

\def\bs{B_s^0}

\def\ks{K_{\sss S}}

\newcommand{\lnp}{H_{NP}^{\Delta B=2}}
\newcommand{\hnp}{H_{eff,NP}^{\Delta B=1}}

\def\barpk{{\raise.35ex\hbox  {${\sss  (}$}}--{\raise.35ex\hbox{${\sss
)}$}}}        \def\bbarp{\hbox{$B$\kern-0.9em\raise1.4ex\hbox{\barpk}}}

   \def\bd{B_d^0} \def\bs{B_s}
   
    \def\ks{K_{\sss    S}}

  \def\rr2{{1\over\sqrt{2}}}

\def\.{\!\cdot\!}    \def\:{\cdots}   \def\[{\left[}   \def\]{\right]}
\def\({\left(} \def\){\right)} 
\begin{document}

\begin{flushright} 
UMISS-HEP-2007-01
\end{flushright}

\begin{center}
\bigskip {\Large  \bf Is there New Physics in B Decays ?\\}  \bigskip   {\large  Alakabha  Datta
$^{a}$\footnote{datta@phy.olemiss.edu} }
\end{center}

\begin{center}
{Department of Physics and Astronomy,}\\
{University of Mississippi,}\\
{Oxford, MS 38677, USA.}\\
\end{center}

\baselineskip=14pt

\begin{center} 
\bigskip (\today) \vskip0.5cm {\Large Abstract\\} \vskip3truemm
\parbox[t]{\textwidth} { 
Rare decays of the $B$ meson are sensitive to new physics effects.
Several experimental results on these decays have been difficult to 
understand within the standard model (SM) though more precise measurements 
and a better understanding of SM theory predictions are needed before 
any firm conclusions can be drawn. In this talk we try to understand the 
present data assuming the presence of new physics. We find that the data 
points to new physics of an extended Higgs sector and we present 
a two higgs doublet model  with a 2-3 flavor symmetry in
 the down type  quark sector that can explain the
 deviations   from  standard   model   reported  in   several  rare   $B$ decays. 
 }
 \end{center}

%

\def\tablefootnote#1{%
\hbox to \textwidth{\hss\vbox{\hsize\captionwidth\footnotesize#1}\hss}} 

\section{Introduction}
Experiments at B factories have demonstrated that the CKM matrix is 
the dominant source of CP violation in hadronic processes specially in decays dominated by tree amplitudes.
These experiments have also confirmed that CP violation in the 
standard model(SM) is large. Now, it is widely expected that 
 new physics (NP) necessary  to stabilize the standard model Higgs mass
 and provide an explanation for electroweak symmetry breaking will be revealed around $\sim $TeV. As CP violation is not a symmetry or an approximate symmetry of nature it is natural to expect this NP to be associated with new CP violating phases which may be large and could be observable in $B$ decays. 
 
 Rare  $B$ decays, where penguin amplitudes play a dominant role, are 
excellent places to look for new physics  CP violating effects \cite{quim}. Decays that 
go through  $b \to s$ transitions are specially interesting as 
 the SM predictions for CP violation in several of these decays are tiny, making them ideal places to look for new physics CP odd phases. It is therefore of crucial importance to test the SM picture of CP violation in these decays.

In fact there are already many measurements in rare  $B$ decays. Interestingly, quite a few of these measurements have been difficult to understand within the SM. First, within the SM, the measurement of
the CP  phase $ \sin {2 \beta}$  in $\bd(t) \to J/\psi  \ks$ should be
approximately  equal to that  in decays  dominated by  the quark-level
penguin transition $ b \to s q{\bar q}$ ($q=u,d,s$) like $\bd(t) \to \phi
K_s$,  $\bd(t) \to \eta^{\prime}  K_s$, $\bd(t)  \to \pi^0  K^0$, etc.
However, there is  a difference between the measurements  of $ \sin {2
\beta}$ in the $ b \to s $ penguin dominated modes ($ \sin {2 \beta} =
0.50 \pm 0.06$) and that in  $\bd(t) \to J/\psi \ks$($ \sin {2 \beta}=
0.685 \pm 0.032)$ \cite{ Babar, Belle,hfag}. Note that the
$ \sin {2 \beta}$ number for the $ b \to s $ penguin dominated modes
is the average of several modes.  The effect of new physics
can be different for different final hadronic states and so the
individual $ \sin {2 \beta}$ measurements for the different modes
are important.
Second, the  latest data on  $B\to\pi K$ decays (branching  ratios and
various  CP  asymmetries)  appear  to  be  inconsistent  with  the  SM
\cite{BKpidecays1,BKpidecays2a, newfit}.  {\footnote{ A cleaner test of the SM
could be provided by looking at the quasi-exclusive decays $B \to K X$
\cite{dattaKX} rather  than the  exclusive $B \to  K \pi$  decays.}  }
Third, within the SM, one  expects no triple-product asymmetries in $B
\to \phi K^*$  \cite{BVVTP}, but BaBar has measured  such an effect at
$1.7\sigma$ level  \cite{BaBarTP}.  There  are  also polarization  anomalies
where the decays $\bd \to \phi K^*$ and $B^+ \to \rho^+ K^*$ appear
to  have large  transverse  polarization amplitudes  in conflict  with
naive SM expectations \cite{hfag,phiKstarexp,rhoKstarexp}.

While  these  deviations certainly  do  not  unambiguously signal  new
physics, they give reason to speculate about NP explanations of the
experimental data.  Furthermore, it is  far more compelling to find NP
scenarios that provide a single solution to all the deviations than to
look  for solutions  to  individual discrepancies.   Taking all  these
deviations seriously one is lead to certain structures of  NP operators
that can  explain the present data\cite{rhoKstar}.
The structure of these operators, which involve scalar-pseudoscalar currents,
 strongly suggest NP associated with an extended Higgs sector.
One of the simplest options is to consider a 2 Higgs doublet model that generates new flavor changing neutral current (FCNC) effects of the right strength. The model that we will consider will generate new FCNC effects at the tree level.
It is known that FCNC $b \to s$ transitions in the SM are not only loop suppressed but are also suppressed by small mixing angles. Hence to produce  effects of the right order( or the same size as the SM FCNC amplitudes) FCNC effects in the NP model must receive additional suppression on top of the suppression due to the mass of the heavy scalar (or pseudoscalar) exchange
which we take to be around a TeV.
The choice for the heavy scalar(pseudoscalar) mass, $m_H$, is based on the assumption that the same new physics needed for the Higgs mass stabilization is also responsible for new FCNC effects in $B$ decays.
 This additional suppression, which is given by $m_s/m_b$ in our model, comes 
from the breaking of a 2-3 flavor symmetry by the strange quark
 mass \cite{model23, bsmixing}. We now give details of the  two Higgs doublet model in the following section.

\section {A Specific NP Model}
Our model of NP is a two Higgs doublet model which has a 2-3
interchange flavor  symmetry in  the down  quark sector  like  the $\mu-\tau$
interchange symmetry in the leptonic  sector \cite{model23, bsmixing}. The
 2-3  symmetry is assumed in the gauge basis where  the mass matrix has
off  diagonal terms and is fully  2-3 symmetric.   Diagonalizing  the mass
matrix  splits  the masses of $s$ and $b$ or  $\mu$ and $\tau$ and  leads to vanishing $m_s(m_\mu)$.
The breaking of the 2-3 symmetry is then introduced though the strange
quark  mass in  the quark  sector and  the muon  mass in  the leptonic
sector.  The  breaking of  the 2-3 symmetry  leads to  flavor changing
neutral currents  (FCNC) in  the quark sector  and the  charged lepton
sector that are suppressed by ${  m_s \over m_b}$ and ${ m_{\mu} \over
m_{\tau}}$ in addition to the mass  of the Higgs boson of
the second Higgs doublet. Additional  FCNC effects of similar size can
be generated  from the  breaking of the  $s-b$ symmetry in  the Yukawa
coupling of the second Higgs doublet. In what follows we will limit our discussions only to the quark sector.

We consider a Lagrangian of the form,  
\beq  {\cal  L}^{Q}_{Y}=  Y^{U}_{ij} \bar  Q_{i,L}  \tilde\phi_1
U_{j,R}  +  Y^D_{ij}  \bar  Q_{i,L}\phi_1 D_{j,R}  +  S^{U}_{ij}  \bar
Q_{i,L}\tilde\phi_2 U_{j,R} +S^D_{ij}\bar Q_{i,L} \phi_2 D_{j,R} \,+\,
h.c. ,
\label{lag1}
\eeq
\noindent where $\phi_i$, for $i=1,2$, are the two scalar doublets of
a 2HDM,  while  $Y^{U,D}_{ij}$ and $S_{ij}^{U,D}$  are the non-diagonal
matrices of the Yukawa couplings. After diagonalizing the $Y$ matrix one can have FCNC couplings associated with the $S$ matrix.

For convenience  we   express $\phi_1$ and  $\phi_2$ in a
suitable basis  such that  only the $Y_{ij}^{U,D}$  couplings generate
the fermion masses. In such a basis one can write,
  \beq 
  \langle\phi_1\rangle=\left(
\begin{array}[]{c}
0\\ {v/\sqrt{2}}
\end{array}
\right)\,\,\,\, , \,\,\,\, \langle\phi_2\rangle=0 \,\,\,.  \eeq
\noindent The two Higgs doublets in this case are of the form,
\bea   
\phi_1   &=  &   \frac{1}{\sqrt{2}}\pmatrix{0   \cr  v+H^0}   +
 \frac{1}{\sqrt{2}}\pmatrix{ \sqrt{2} \chi^+ \cr i \chi^0}, \nonumber\\
 \phi_2 &= & \frac{1}{\sqrt{2}}\pmatrix{ \sqrt{2} H^+ \cr H^1+i H^2}. \
 \eea
 
In principle there  can be mixing among the neutral  Higgs but here we
neglect such mixing.
We assume  the doublet $\phi_1$  corresponds to the scalar  doublet of
the SM and  $H^0$ to the SM Higgs field.  In  addition, we assume that
the second  Higgs doublet does  not couple to the  up-type quarks($S^U
\equiv 0$). For the down type couplings in Eq.~\ref{lag1} we have,
\beq   {\cal  L}^{D}_{Y}=  Y^D_{ij}   \bar  Q_{i,L}\phi_1   D_{j,R}  +
S^D_{ij}\bar Q_{i,L} \phi_2 D_{j,R} \,+\, h.c.
\label{lag2}
\eeq We  assume the  following symmetries for  the matrices  $Y^D$ and
$S^D$:
\begin{itemize}

\item{ There  is a discrete symmetry under  which $d_{L,R} \rightarrow
-d_{L,R}$}

\item{ There is a $s-b$ interchange symmetry: $s \leftrightarrow b$}

\end{itemize}

The discrete symmetry involving the  down quark is enforced to prevent
$ s \to d $ transition because of constraints from the kaon system. It
also prevents  $ b \to d $  transitions since $B_d$ mixing  as well as
the value  of $ \sin { 2  \beta}$ measured in $\bd(t)  \to J/\psi \ks$
are consistent with  SM predictions. Although there may  still be room
for NP  in $ b \to d$  transitions, almost all deviations  from the SM have
been reported only in $ b \to s$ transitions and so we assume no NP in
$ b \to d$ transitions in this work.

The above symmetries then give  the following structure for the Yukawa
matrices,
\begin{eqnarray}
 Y^D &= &\pmatrix{y_{11} & 0 & 0  \cr 0 & y_{22} & y_{23}\cr 0 &y_{23}
& y_{22}}, \nonumber\\ 
S^D &= &\pmatrix{s_{11} & 0 & 0 \cr 0 & s_{22} &
s_{23}\cr 0 &s_{23} & s_{22}}. \
 \label{Yukawas}
\end{eqnarray}

The diagonalizing of $Y^D$ reduces $S^D$ to a diagonal form also and so there are no FCNC effects associated with the second Higgs doublet. To introduce FCNC effects we
 introduce  the strange quark  mass as a
small breaking of the $s-b$ symmetry and consider the structure,
\begin{eqnarray}
 Y^D_n &= &\pmatrix{y_{11} & 0 &  0 \cr 0 & y_{22}(1+2z) & y_{22}\cr 0
&y_{22} & y_{22}}, \
 \label{symbreak}
 \eea
with $z  \sim 2  m_s/m_b$ being a  small number.

At this point we can consider two scenarios: the first   corresponds to the situation where the matrix $S^D$ is still $s-b$ symmetric. This leads to an interesting prediction that there are no observable new weak phase in $\bs$ mixing \cite{bsmixing}.  In the second case we will allow for small breaking of the $s-b$ symmetry in $S^D$  in Eq.~\ref{Yukawas}. This  then results in   an observable phase in $\bs$ mixing. In this talk we will consider only the first scenario.

{}For the first case,
   $S^D$  in the mass  eigenstate basis  has
the form,
\bea
S^D  \rightarrow  S^{D'} &= &\pmatrix{s_{d}e^{ i \phi_{dd}} & 0 &
0 \cr
0
& s_{s}e^{ i \phi_{ss}} & z s e^{ i \phi_{sb}} \cr
0
&z s e^{ i \phi_{sb}} &
 s_{b} e^{ i \phi_{bb}} }, \
 \label{sd_case_a}
\end{eqnarray}
where
 \bea
s_{d}e^{ i \phi_{dd}} & = & s_{11}, \nonumber\\
s_{s}e^{ i \phi_{ss}} & =& (s_{22}-s_{23} ), \nonumber\\
s_{b}e^{ i \phi_{bb}} & = & (s_{22}+s_{23} ),   \nonumber\\ 
{s}e^{ i \phi_{sb}} & = & s_{23}. \ 
\eea
 After integrating out the heavy Higgs bosons, $H^{1,2}$, which we shall henceforth rename $S$ and $P$ bosons,  we can generate the following effective
Hamiltonian for $\Delta B=2$ and $\Delta B=1$ processes,
\bea
\lnp & = & \frac{1}{2 {m^2_{S(P)}}}z^2 s^2
\left[\pm e^{ i 2 \phi_{sb}}O_{RR}  \pm e^{ -i 2 \phi_{sb}}O_{LL} +O_{RL} +O_{LR} \right], \
\label{mixing}
\eea
with
\bea
O_{LL} & = &  \bar{s} ( 1-\gamma_5) b \bar{s} (1- \gamma_5) b, \nonumber\\
O_{LR} & = &  \bar{s} ( 1-\gamma_5) b \bar{s} (1+ \gamma_5) b, \nonumber\\
O_{RL} & = &  \bar{s}( 1+\gamma_5) b \bar{s} (1- \gamma_5) b, \nonumber\\
O_{RR} & = &  \bar{s} ( 1+\gamma_5) b \bar{s}(1+ \gamma_5) b, \
\label{vector-scalar}
\eea
and
\bea
\hnp &= & 4\frac{G_F}{\sqrt{2}}z \frac{s s_{q}}{g^2}
\frac{m_W^2}{{m^2_{S(P)}}}
\left[\pm e^{ i  \phi_{+}}H_{RR}  \pm e^{- i  \phi_{+}}H_{LL}\right. \nonumber\\
&  & \left. +e^{ i  \phi_{-}}H_{RL} +e^{ -i  \phi_{-}}H_{LR} \right], \
\label{heff}
\eea
 where the $\pm$ sign is for $S$ and $P$ exchange interaction, $\phi_{\pm}= \phi_{sb} \pm \phi_{qq}$ and
 \bea
H_{LL} & = &  \bar{s} ( 1-\gamma_5) b \bar{q} (1- \gamma_5) q, \nonumber\\
H_{RL} & = &  \bar{s}( 1+\gamma_5) b \bar{q} (1- \gamma_5) q, \nonumber\\
H_{LR} & = &  \bar{s} ( 1-\gamma_5) b \bar{q} (1+ \gamma_5) q,\nonumber\\
 H_{RR} & = &  \bar{s} ( 1+\gamma_5) b \bar{q} (1+ \gamma_5) q. \
\label{heffgeneral}
\eea

One can now study the predictions of this model for various FCNC processes involving the $B$ meson. The structure of $\hnp$ has just the right form found in Ref~\cite{rhoKstar} to explain the various discrepancies in the rare $B$ decays observed in experiments.
In the next section we briefly discuss the predictions of the model.

\section{Predictions and Conclusions} 
The predictions of our model are given below:
 \begin{itemize}
 
 \item{
 As mentioned earlier, the effective $ \sin {2 \beta}$ measured in pure penguin or penguin dominated $ b \to s $ transitions are generally lower than
 that from $\sin{2 \beta}$ measured in $ \bd \to J/\psi K_s$.
 This is now a problem which is almost four years old and has not been resolved yet \cite{dattarparity}. In our model
 the effective $ \sin {2 \beta}$ measured in $ b \to s $ transitions like 
$\bd(t) \to \phi K_s$, $\bd(t) \to \eta(\eta') K_s$, $ \bd(t) \to K_s \pi^0$ etc are different and smaller than from $\sin{2 \beta}$ measured in $ \bd \to J/\psi K_s$. This is consistent with present experiments.
Large branching ratios in charmless semi-inclusive $B \to \eta' X_s$ decays also have been difficult to understand in the SM \cite{dattaetaprime} and it is plausible that our NP model may provide an explanation of this large 
branching ratio.
}

\item{
The model gives a good fit to the data on the various $ \bd \to K \pi$ data \cite{rosner}.
}

\item
 { The model explains the large transverse polarizations observed in $\bd \to \phi K^*$ and $B^+ \to K^{*0} \rho^+$ but not in
 $B^+ \to K^{*+} \rho^0$ \cite{pol}
 }
 
 \item {The predictions for $\bs$ mixing \cite{bsmixing} are consistent with experiments \cite{D0CDF}. The phase in $\bs$ mixing may or may not be observable even though new weak phases will be observable in hadronic $ b \to s \bar{q}{q}$ transitions where $q=u, d,s$.
 }
 
 \end{itemize}
 In conclusion, we expect NP to be exist around a scale of $TeV$ to stabilize 
the Higgs mass and provide an explanation for electroweak symmetry breaking. It is quite possible that this NP will contain new CP violating phases and since CP is not a symmetry of nature these CP odd phases can be large and hence detectable in rare $B$ decays. Interestingly there
  are now several $B$ decay modes in which there
 appear  to be deviations from the SM predictions. These deviations
 could signal the presence of beyond the SM physics.  In this work we
 were interested in a NP model that provides an explanation of
  all the deviations seen in several rare $B$ decays.  We considered a two Higgs doublet model with
 a 2-3 symmetry in the down type quark sector.
 The breaking of the 2-3 symmetry, introduced by the strange quark
 mass leads to FCNC in the quark sector 
  that are suppressed by ${ m_s \over m_b}$ 
  in addition to the mass of the heavy
 Higgs boson which is taken to be around a TeV. This model can explain all the deviations so far reported in several rare $B$ decays and future precise measurements at $B$ factories as well as results from upcoming collider experiments such as LHC will be able to test this model of new physics.

\end{document}